\documentclass[preprint,showpacs,preprintnumbers,amsmath,amssymb,nofootinbib,showkeys,aps]{revtex4}
\pdfoutput=1
\usepackage{epsfig}

\usepackage{graphicx}
\usepackage{dcolumn}
\usepackage{bm}
\usepackage{amsmath}
\usepackage{amssymb}
\usepackage{latexsym}
\usepackage{color}
\usepackage{hyperref}
\usepackage{mathrsfs}

\newcommand{\be}{\begin{equation}}
\newcommand{\ee}{\end{equation}}
\newcommand{\bea}{\begin{eqnarray}}
\newcommand{\eea}{\end{eqnarray}}
\newcommand{\texpdf}{\texorpdfstring}

\begin{document}


\title{A Method for Obtaining Cosmological Models Consistency Relations and Gaussian Processes Testing}

\author{J. F. Jesus$^{1,2}$}\email{jf.jesus@unesp.br}
\author{A. A. Escobal$^{2}$}\email{anderson.aescobal@gmail.com}
\author{R. Valentim$^{3}$}\email{valentim.rodolfo@unifesp.br}
\author{S. H. Pereira$^{2}$}\email{s.pereira@unesp.br}

\affiliation{$^1$Universidade Estadual Paulista (UNESP), Campus Experimental de Itapeva - R. Geraldo Alckmin, 519, 18409-010, Itapeva, SP, Brazil,
\\$^2$Universidade Estadual Paulista (UNESP), Faculdade de Engenharia e Ci\^encias de Guaratinguet\'a, Departamento de F\'isica - Av. Dr. Ariberto Pereira da Cunha 333, 12516-410, Guaratinguet\'a, SP, Brazil
\\$^3$Universidade Federal de S\~ao Paulo (UNIFESP), Departamento de F\'{\i}sica, Instituto de Ci\^encias Ambientais, Qu\'{\i}micas e Farmac\^euticas (ICAQF), Rua S\~ao Nicolau 210, 09913-030, Diadema, SP, Brazil}


\def\zt{\mbox{$z_t$}}

\begin{abstract}

In the present work, we apply consistency relation tests to several cosmological models, including the flat and non-flat $\Lambda$CDM models, as well as the flat XCDM model. The analysis uses a non-parametric Gaussian Processes method to reconstruct various cosmological quantities of interest, such as the Hubble parameter $H(z)$ and its derivatives from $H(z)$ data, as well as the comoving distance and its derivatives from SNe Ia data. We construct consistency relations from these quantities which should be valid only in the context of each model and test them with the current data. We were able to find a general method of constructing such consistency relations in the context of $H(z)$ reconstruction. In the case of comoving distance reconstruction, there were not a general method of constructing such relations and this work had to write an specific consistency relation for each model. From $H(z)$ data, we have analyzed consistency relations for all the three above mentioned models, while for SNe Ia data we have analyzed consistency relations only for flat and non-flat $\Lambda$CDM models. Concerning the flat $\Lambda$CDM model, some inconsistency was found, at more than $2\sigma$ c.l., with the $H(z)$ data in the interval $1.8\lesssim z\lesssim2.4$, while the other models were all consistent at this c.l. Concerning the SNe Ia data, the flat $\Lambda$CDM model was consistent in the $0<z<2.5$ interval, at $1\sigma$ c.l., while the nonflat $\Lambda$CDM model was consistent in the same interval, at 2$\sigma$ c.l.
\end{abstract}

\maketitle



\section{Introduction} 
It is well known that the standard model of cosmology, called flat $\Lambda$CDM model or just concordance model, is consistent with most observations to date, both at background level as well as at linear perturbations about a spatially homogeneous and isotropic metric. The model correctly probe a vast range of scales in space and time by using just a small set of parameters. A key role is played by the cosmological constant term $\Lambda$, which correctly explains the recent acceleration of the universe. Additionally, a power-law form of the primordial scalar perturbation spectrum agrees well with most observational data sets, including the distance-redshift relation of type Ia supernovae (SNe Ia), Cosmic Microwave Background (CMB) data, Hubble parameter ($H(z)$) data, Baryon Acoustic Oscillation (BAO) data, weak lensing shear and data from large scale structure.

Despite of its success, some tension has been noticed recently while fitting the concordance model to different data sets. One of them is known as the Hubble tension, regarding the $4\,\sigma$ discrepancy between the value obtained for the Hubble parameter $H_0$ when derived from local observations as compared to CMB and BAO data. This and other limitations of the flat $\Lambda$CDM model (see \citep{P2014,Riess:
2020sih,Martinelli:2019krf,Bull:2015stt} for detailed discussions about that) gave rise to several alternative models to try to solve the $\Lambda$CDM problems. Parallel to this increase in the number of alternative models, methods to test the consistency of different models were also developed, and this is the subject that this work wants to explore.

The first consistency tests were developed to compare the standard model against time-varying dark energy models. The Statefinder hierarchy method was introduced as a geometrical diagnostic of dark energy \cite{Sahni:2002fz,Arabsalmani:2011fz}, being algebraically related
to the equation of state of dark energy and its first time derivative. Sahni, Shafieloo and Starobinsky \citep{SahniEtAl08} introduced two diagnostic methods to test dark energy models. The first, named $Om(z)$, is a combination of the cosmological redshift and the Hubble parameter. Basically, if the value of $Om(z)$ is constant for different redshifts, then the dark energy corresponds exactly to the cosmological constant term. The second diagnostic, named \emph{acceleration probe $\bar{q}$}, can be used to determine the cosmological redshift at which the universe began to accelerate. The method calculates the mean value of the deceleration parameter over a small redshift
range. Both diagnostic methods are performed without reference to the current value of the matter density. Then, the same authors introduced the $Om3$ diagnostic \cite{Shafieloo:2012rs}, which combines standard
candle information from SNe Ia with standard ruler information from BAO to yield a novel null diagnostic of the
cosmological constant paradigm.  The test was recently used with the most recent BAO observables from the eBOSS survey \cite{Shafieloo:2022gbw}. Other tests were also developed. In \cite{Shafieloo:2009hi}, Shafieloo and Clarkson discussed how direct measurements of the Hubble parameter can be used to test the
standard paradigm in cosmology. Particularly, such measurements can be used to search for deviations from
standard model, being independent of
dark energy or a metric based theory of gravity. Null tests using supernova data were done in \cite{Yahya:2013xma} by using Gaussian Processes (GP) in order to take model-independent derivatives of the distance. They found that, by using the Union 2.1 data, the concordance
model is compatible but the error bars are large. When simulated data sets were used for the Dark Energy Survey, a sharper null test of the cosmological constant were obtained. Similar consistency tests for the cosmological constant $\Lambda$, for the Copernican principle and for different dark energy models were developed earlier \cite{Zunckel:2008ti,Clarkson:2007pz,Clarkson:2012bg}. 

In the present work, we check the consistency of the flat and non-flat $\Lambda$CDM models and also the flat XCDM model. We have extended the analysis of \cite{Yahya:2013xma} since that, in addition to SNe Ia data, we have also added $H(z)$ data, as well as have used more recent SNe Ia data, namely, the Pantheon$+$ compilation. Non-parametric GP method was used to reconstruct cosmological quantities of interest, as the Hubble parameter $H(z)$ and its derivatives and also comoving distance and its derivatives and we have constructed consistency relations from these quantities.

The paper is organized as follows. The equations of the model are presented in Section~\ref{basic}. The cosmological data sample used and methodology are in Section~\ref{datamet}. Section~\ref{analysis} describes our analysis and results. Conclusions are in Section~\ref{conclusion}. 

\section{\label{basic}Cosmological equations}
To evaluate the consistency of the most accepted cosmological models in the context of the FLRW metric, we will perform tests using data from the late universe at the background level.
In order to do that, we will adopt non-parametric methods to reconstruct quantities of interest in each model. These quantities will be defined below.
\subsection{Flat \texpdf{$\Lambda$}{L}CDM}
\subsubsection{Consistency relations from $H(z)$ data}
The Friedmann equation in the FLRW metric relates the Hubble parameter $H(z)$ with the material content of the universe. For the spatially flat $\Lambda$CDM model, it is written as:
\be
H(z)^2=H_0^2 \left[\Omega_m(1+z)^3+1-\Omega_m\right]\,.
\label{eqH}
\ee
By deriving it with respect to the redshift:
\be
2 H(z) H'(z)=3H_0^2\Omega_m(1+z)^2\,,
\label{eqdH}
\ee
where the prime denotes derivative with respect to redshift. Deriving again:
\be
2 H(z) H''(z)+2 H'(z)^2=6H_0^2\Omega_m(1+z)\,.
\label{eqd2H}
\ee

The eqs. \eqref{eqH}-\eqref{eqdH} form a system which can be solved to furnish the $H_0^2$ and $\Omega_m$ parameters as:
\begin{align}
H_0^2= \frac{H(z) \left[3 (z+1)^2 H(z)-2 z \left(z^2+3 z+3\right)H'(z)\right]}{3 (1+z)^2}\,,\label{H02}\\
   \Omega_m= -\frac{2 H'(z)}{2 z \left(z^2+3 z+3\right)
   H'(z)-3 (1+z)^2 H(z)}\,.\label{wm}
\end{align}

If now we use Eqs. \eqref{H02} and \eqref{wm} to eliminate $H_0$ and $\Omega_m$ on Eq. \eqref{eqd2H}, we can find the relation:
\be
H(z) H''(z)+H'(z)^2=\frac{2 H(z) H'(z)}{1+z}\,,
\ee
or, if we solve for $H''(z)$:
\be\label{H2FLCDM}
H''(z)=\frac{2H'(z)}{1+z}-\frac{H'(z)^2}{H(z)}\,.
\ee

We may see this equation as a prediction from the flat $\Lambda$CDM model. One may reconstruct $H''(z)$ from $H(z)$ data and compare it with the reconstruction of the rhs of \eqref{H2FLCDM}. Instead of comparing two reconstructions, let us define and reconstruct the following diagnostic $\delta_{H}^{(2)}$:
\be\label{dHFLCDM}
\delta_{H{F\Lambda \text{CDM}}}^{(2)}\equiv H''_{GP}-H''_{model}\,,
\ee
where $H''_{GP}$ is the $H''(z)$ obtained from data using the Gaussian Processes and $H''_{model}$ is the rhs of \eqref{H2FLCDM} (the model prediction for $H''(z)$).

\subsubsection{Consistency relations from SNe Ia}
Since that SNe Ia data do not constrain $H_0$, one has to work with $E(z)\equiv\frac{H(z)}{H_0}$. We have, for flat $\Lambda$CDM:
\be
E(z)^2=\Omega_m(1+z)^3+1-\Omega_m\,.
\ee

From this relation we can find for $\Omega_m$:
\be
\Omega_m=\frac{E(z)^2-1}{(1+z)^3-1}\,.
\label{Omz}
\ee
This corresponds to the $Om(z)$ diagnostics \cite{SahniEtAl08}. Assuming a spatially flat Universe, we can relate this to the comoving distance $D_C(z)$:
\be
D_C(z)=\int_0^z\frac{dz'}{E(z')}\,,
\ee
so that $E(z)=\frac{1}{D_C'(z)}$. From (\ref{Omz}) we can write:
\be\label{wmdcflcdm}
\Omega_m=\frac{1}{(1+z)^3-1}\left(\frac{1}{D_C'(z)^2}-1\right)\,.
\ee
Deriving the expression \eqref{wmdcflcdm} and isolating the largest derivative of $D_C(z)$, we get:
\be
D_C''(z)=\frac{3(1+z)^2 D_C'(z) \left(D_C'(z)^2-1\right)}{2 z \left(z^2+3 z+3\right)}\,.
\label{DC2FL}
\ee

As one may see from Eq. \eqref{DC2FL}, it may appear a divergence for $D_C''$ as $z$ tends to zero, so the GP method may not reconstruct well $D_C''(z)$ in this case, as it does not work well with rapidly varying functions \cite{Seikel:2012uu}. So, in order to test the validity of flat $\Lambda$CDM model against SNe Ia data, we shall define the following function:
\be\label{LFLCDM}
\mathcal{L}_{F\Lambda \text{CDM}}\equiv 2 z \left(z^2+3 z+3\right)D_C''(z)-3(1+z)^2 D_C'(z) \left(D_C'(z)^2-1\right)\,,
\ee
which corresponds to the $\mathcal{L}$ defined by \cite{Zunckel:2008ti}.




\subsection{Nonflat \texpdf{$\Lambda$}{L}CDM}
\subsubsection{Consistency relations from $H(z)$ data}
For the nonflat $\Lambda$CDM model, the Friedmann equation is
\be
H(z)^2=H_0^2\left[\Omega_m(1+z)^3 + (1-\Omega_\Lambda-\Omega_m)(1+z)^2 + \Omega_\Lambda\right]\,,
\label{Hnfl}
\ee
where the curvature parameter is given by $\Omega_k=1-\Omega_\Lambda-\Omega_m$. By deriving \eqref{Hnfl}:
\be
2 H(z) H'(z)=H_0^2 \left[2(1+z) (1-\Omega_\Lambda-\Omega_m)+3\Omega_m(1+z)^2\right]\,,
\label{Hnfl1}
\ee
and:
\be
H(z) H''(z)+H'(z)^2=H_0^2\left[(1-\Omega_\Lambda-\Omega_m)+3\Omega_m (1+z)\right]\,.
\label{Hnfl2}
\ee
Deriving again:
\be
H(z)H'''(z)+3H'(z)H''(z)=3H_0^2\Omega_m\,.
\label{H3LCDM}
\ee

By solving Eqs. \eqref{Hnfl}-\eqref{Hnfl2} to obtain parameters $H_0$, $\Omega_m$ and $\Omega_\Lambda$ and inserting the results in \eqref{H3LCDM}, we can find:
\be
2\left[H(z) H''(z)+H'(z)^2\right]=H(z) \left[(1+z) H'''(z)+\frac{2 H'(z)}{1+z}\right]+3 (1+z) H'(z) H''(z)\,,
\ee
or, solving for $H'''(z)$:
\be\label{d3H}
H'''(z)=\frac{(1+z) H'(z) \left[2 H'(z)-3 (1+z) H''(z)\right]+H(z) \left[2 (1+z) H''(z)-2 H'(z)\right]}{(1+z)^2 H(z)}\,.
\ee

We define $H'''_{model}\equiv H'''(z)$, where $H'''(z)$ is given by the rhs of the expression \eqref{d3H} above. So, we define a consistency relation $\delta_{H}^{(3)}$ for the $\Lambda$CDM model based on the $H(z)$ data, as given by:
\begin{align}\label{dHLCDM}
    \delta_{H{\Lambda \text{CDM}}}^{(3)} \equiv H'''_{GP}-H'''_{model}\,.
\end{align}

Where $H'''_{GP}$ is the reconstruction of the third derivative of $H(z)$ obtained by GP.
\subsubsection{Consistency relations from SNe Ia}
For SNe Ia, we have the relation:
\be\label{EDM}
E(z)^2=\frac{1+\Omega_kD_M(z)^2}{D_M'(z)^2}\,,
\ee
where $D_M(z)$ is dimensionless transverse comoving distance \cite{Hogg99}. From \eqref{Hnfl}, we may write:
\be
E(z)^2=\Omega_m(1+z)^3 + \Omega_k(1+z)^2 + 1-\Omega_m-\Omega_k\,,
\label{Enfl}
\ee
where we have used the normalization condition $\Omega_m+\Omega_\Lambda+\Omega_k=1$. Solving \eqref{Enfl} for $\Omega_m$, we have:
\be
\Omega_m=\frac{E(z)^2-\Omega_k z^2-2 \Omega_k z-1}{z \left(z^2+3 z+3\right)}\,.
\label{wmwkEz}
\ee

Using \eqref{EDM} in order to write $\Omega_m$ in terms of $D_M$, we have
\be
\Omega_m=\frac{\Omega_k D_M(z)^2+1-D_M'(z)^2\left[\Omega_k z (2+z)+1\right]}{z(z^2+3 z+3)D_M'(z)^2}\,.
\ee

Deriving it with respect to redshift, and solving for $\Omega_k$, in compact form we get
\be
\Omega_k=\frac{2 z \left(z^2+3 z+3\right) D_M''-3 (1+z)^2 D_M'^3+3 (1+z)^2 D_M'}{z^2 \left(z^2+4 z+3\right) D_M'^3+2 z
   \left(z^2+3 z+3\right) D_M D_M'^2-D_M^2 \left(2 z \left(z^2+3 z+3\right) D_M''+3 (1+z)^2 D_M'\right)}\,.
\ee

Deriving it again and solving for $D_M'''$, we find:
\begin{align}
    D_M'''(z) = \dfrac{g(z)}{f(z)}\,,
\end{align}
where $g(z)$ is given by:
\begin{align}
g(z)\equiv&\, 3 (1+z) \left(z^2 (3+z) D_M''(z)^2-2 (1+z) D_M'(z)^4+2 (1+z) D_M'(z)^2+\right.\nonumber\\
&+2 z (2+z) D_M'(z) D_M''(z)+D_M(z)
   \left(D_M'(z)^2-1\right) \left(3 (1+z) D_M''(z)+2 D_M'(z)\right)\nonumber\\
&\left.-D_M(z)^2 D_M''(z) \left(3 (1+z) D_M''(z)+2
   D_M'(z)\right)\right)\,,
   \label{fDM3}
\end{align}
and $f(z)$ is
\be
f(z)\equiv\left(z^2+4 z+3\right) z^2 D_M'(z)-3 (1+z)^2 D_M(z)^2 D_M'(z)+2 \left(z^2+3 z+3\right) z D_M(z)\,.
\ee

For the nonflat $\Lambda$CDM model, we define the consistency relation $\mathcal{L}_{\Lambda CDM}$, as
\be\label{LLCDM}
\mathcal{L}_{\Lambda \text{CDM}}\equiv f(z)D_{M,GP}'''-g(z)\,,
\ee
where $D_{M,GP}'''$ is the reconstruction of the third derivative of $D_M(z)$ obtained by GP.



\subsection{Flat XCDM}
\subsubsection{Consistency relations from $H(z)$ data}
In the flat XCDM model the Friedmann equation is
\be
H(z)^2=H_0^2 \left[\Omega_m (1+z)^3+(1-\Omega_m) (1+z)^{3 (w+1)}\right]\,.
\label{Hx}
\ee
Carrying out the first, second and third derivatives, respectively, we get
\begin{align}
2 H(z) H'(z)&=3 H_0^2 (1+z)^2 \left[\Omega_m+(w+1)(1-\Omega_m) (1+z)^{3 w}\right]\label{dHx}\,,\\
2 \left(H(z) H''(z)+H'(z)^2\right)&=3 H_0^2 (1+z) \left[2\Omega_m+(w+1)(3w+2) (1-\Omega_m) (1+z)^{3 w}\right]\label{d2Hx}\,,\\\label{d3HFXCDM}
2 H(z) H'''(z)+6 H'(z) H''(z)&=H_0^2 \left[6\Omega_m+3(w+1) (3 w+1) (3 w+2) (1-\Omega_m) (1+z)^{3 w}\right]\,.
\end{align}
Solving the \eqref{Hx}-\eqref{d3HFXCDM} and isolating the highest order derivative of $H(z)$, we obtain
\begin{align}\nonumber
H'''(z)=& - \frac{3H'(z) H''(z)}{2H(z)}+\frac{H(z)\left[(1+z)^2 H''(z)^2+6 H'(z)^2-4 (1+z) H'(z) H''(z)\right]}{(1+z)H(z) \left[2(1+z) H'(z)-3 H(z)\right]}+\\\label{d3HFX}
&+\frac{(1+z)^3 H'(z)^4-3 H(z)^3H'(z)+2 
   (1+z)^2 H(z) H'(z)^2 \left((1+z) H''(z)-2 H'(z)\right)}{(1+z)^2H(z)^2\left[2(1+z) H'(z)-3 H(z)\right]}\,.
\end{align}

We define $H'''_{model}\equiv H'''(z)$, where $H'''(z)$ is given by \eqref{d3HFX}. So the consistency relation $\delta_{H}^{(3)}$ for the XCDM model based on the $H(z)$ data is given by:
\begin{align}\label{DHFXCDM}
    \delta_{H{F\text{XCDM}}}^{(3)} \equiv H'''_{GP}-H'''_{model}\,.
\end{align}

\subsubsection{Consistency relations from SNe Ia}
It was not possible to calculate consistency relations from SNe Ia for flat XCDM because it was not possible to solve analytically for $(w,\Omega_m,H_0)$ from $(D_C(z),D_C'(z),D_C''(z))$ equations. One should resort to numerical methods for solving a nonlinear system of equations, which is beyond the scope of the present work.

\section{\label{datamet}Dataset and methodology}
The observational datasets used in this work is composed of SNe Ia data from the Pantheon$+$ sample \cite{pantheonplus} and a compilation of measurements of the Hubble parameter $H(z)$ estimated from Cosmic Chronometers \cite{{Moresco22}}.

We work with $32$ astrophysical measurements of $H(z)$ with covariance,
obtained by estimating the differential ages of galaxies \cite{{Moresco22}}, called Cosmic Chronometers, in the redshift range $0.07<z<1.97$. The Pantheon$+$ data sample \cite{pantheonplus}, consists of 1701 light curves of $1550$ distinct SNe Ia, in the redshift range $0.001<z<2.26$.

We use the non-parametric method called Gaussian Processes (GP) \cite{Seikel:2012uu}, to reconstruct the Hubble parameter $H(z)$, from cosmic timer data, and the transverse moving distance $D_M(z) $ of the Pantheon$+$ sample data. To obtain the reconstructions with the GPs, it is necessary to define a correlation function (kernel)\cite{Jesus:2019nnk}, $k(x_i,x_j)$, between the points $x_i$ and $x_j$ of the data sample. In this work, we use the Exponential Square kernel given by:
\begin{align}
k(x_i,x_j)=\sigma_f^2\exp\left[-\frac{(x_i-x_j)^2}{2\ell^2}\right]\,,
\end{align}
where $\ell$ and $\sigma_f$ are the GP hyperparameters that are obtained from the data.

\begin{figure}[h]
\begin{center}
\includegraphics[width=.49\textwidth]{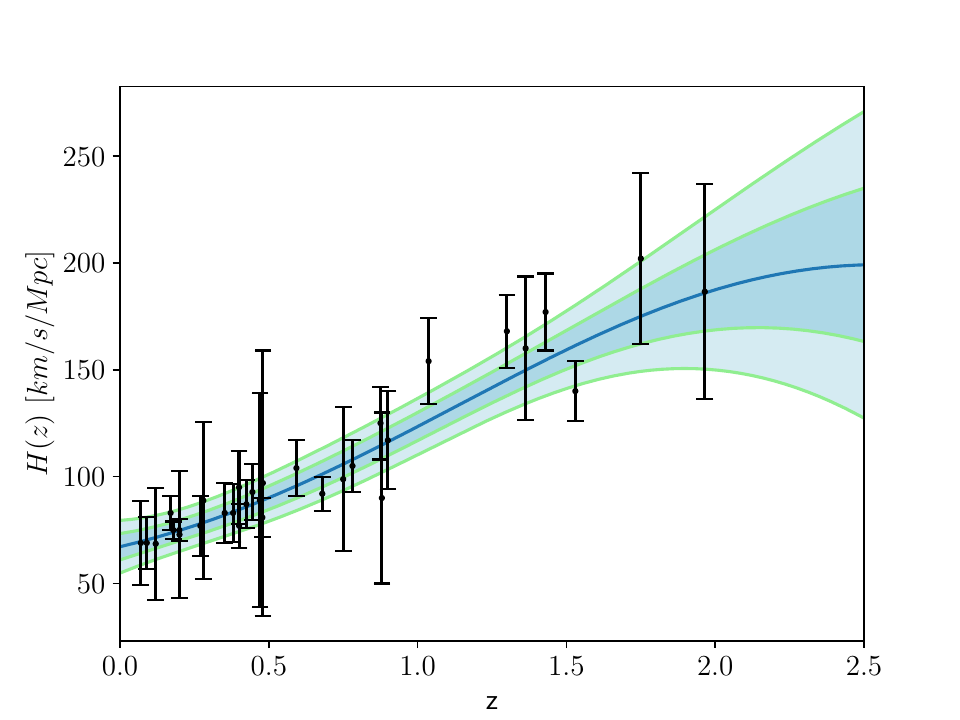}
\includegraphics[width=.49\textwidth]{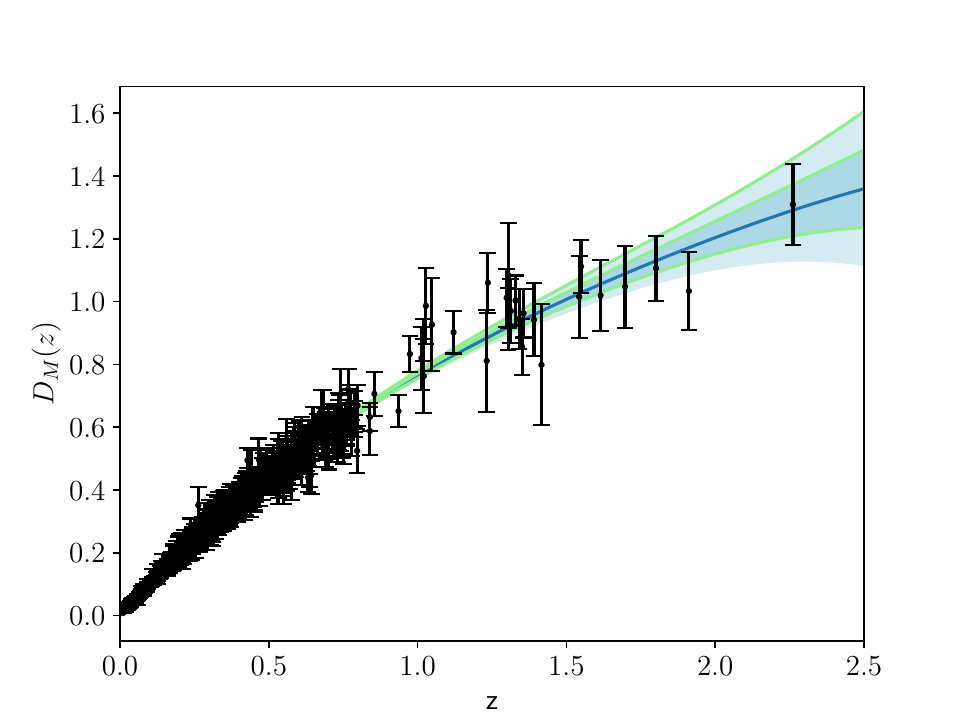}
\end{center}
\caption{GP reconstructions from data. \textbf{a) Left:} Reconstruction of $H(z)$ (in km/s/Mpc) function from 32 $H(z)$ data with covariance. \textbf{b) Right:} Reconstruction of dimensionless $D_M(z)$ from Pantheon$+$ SNe Ia data.}
\label{recdados}
\end{figure}

These reconstructions are shown in Fig. \ref{recdados}, where we also present the observational data of each sample.

GP allows reconstructing not only the functions $H(z)$ and $D_M(z)$, but also their derivatives. And with that, we can then reconstruct the consistency relations discussed above.

\section{\label{analysis}Analyses and Results}
To perform consistency relations tests based on observational data, the non-parametric GP method was used to reconstruct cosmological quantities of interest from the available data sample. The Hubble parameter $H(z)$ and its derivatives were reconstructed from the cosmic chronometers data, while the transverse comoving distance $D_M(z)$ and its derivatives were reconstructed from the Pantheon$+$ sample data. With these reconstructed quantities, it was possible to test the consistency relations for each cosmological model considered, using only the available data samples.

\subsection{Consistency tests for the $H(z)$ data}
\begin{figure}[!h]
    \centering
    \includegraphics[width=0.8\textwidth]{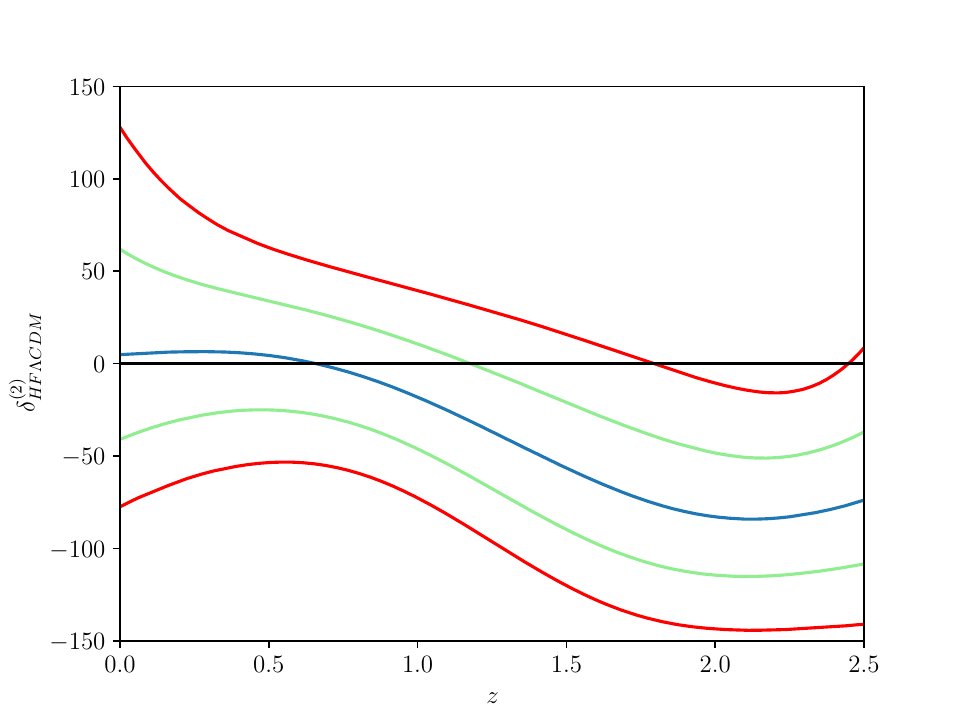}
     \caption{Plot of $\delta^{(2)}_{HF\Lambda CDM}=H''_{GP}-H''_{F\Lambda \text{CDM}}$, the difference between the Hubble parameter second derivative obtained from GP reconstruction and the same quantity obtained from the flat $\Lambda$CDM relation, as explained on the text.}
     \label{pdFLCDM}
\end{figure}

From the reconstruction of the $H(z)$ function and its derivatives using the GP, it was possible to carry out consistency tests in the analyzed models. Fig. \ref{pdFLCDM} presents the reconstruction of $\delta_{H{F\Lambda \text{CDM}}}^{(2)}$ for the flat $\Lambda$CDM model, which is given by the expression \eqref{dHFLCDM}. The reconstruction is compatible with zero at 2$\sigma$ c.l. in almost all the analyzed redshift range, except in the interval of $1.79\lesssim z\lesssim2.45$, in which the reconstruction becomes negative in more than 2$\sigma$.

\begin{figure}[!h]
    \centering
    \includegraphics[width=0.8\textwidth]{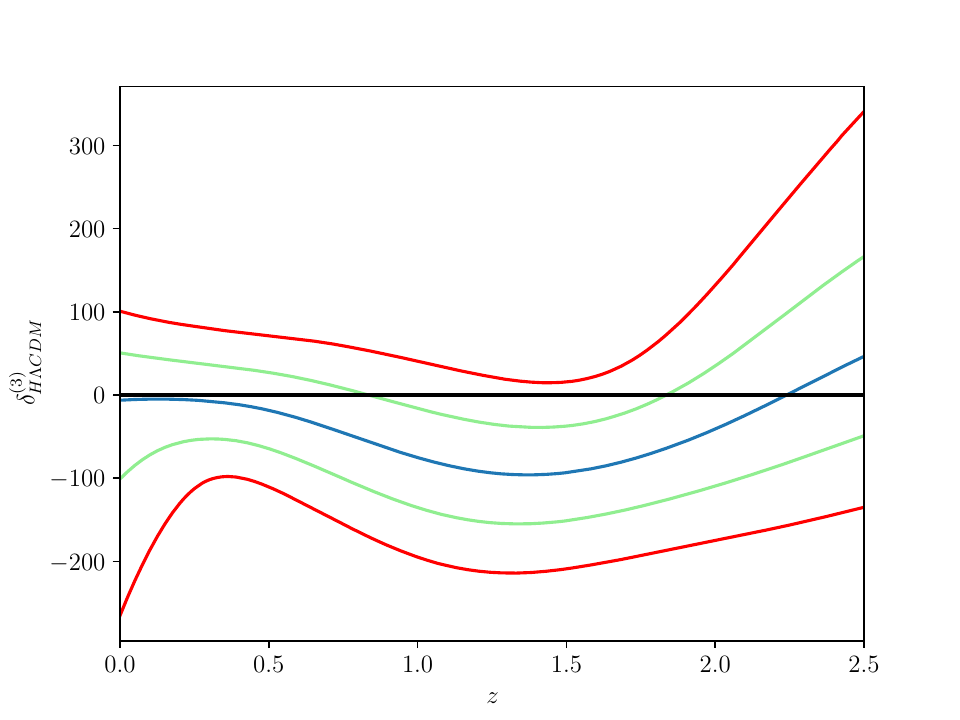}
     \caption{Plot of $\delta^{(3)}_{H\Lambda \text{CDM}}=H'''_{GP}-H'''_{\Lambda \text{CDM}}$, the difference between the Hubble parameter third derivative obtained from GP reconstruction and the same quantity obtained from the nonflat $\Lambda$CDM relation, as explained on the text.}
     \label{pdLCDM}
\end{figure}

The $\delta^{(3)}_{H{\Lambda \text{CDM}}}$ for the $\Lambda$CDM model, given by the relation \eqref{dHLCDM} and reconstructed by GP, is shown in Fig. \ref{pdLCDM}. We obtain a compatibility with zero within an uncertainty interval of 2$\sigma$ in the entire redshift range analyzed.

\begin{figure}[!h]
    \centering
    \includegraphics[width=0.8\textwidth]{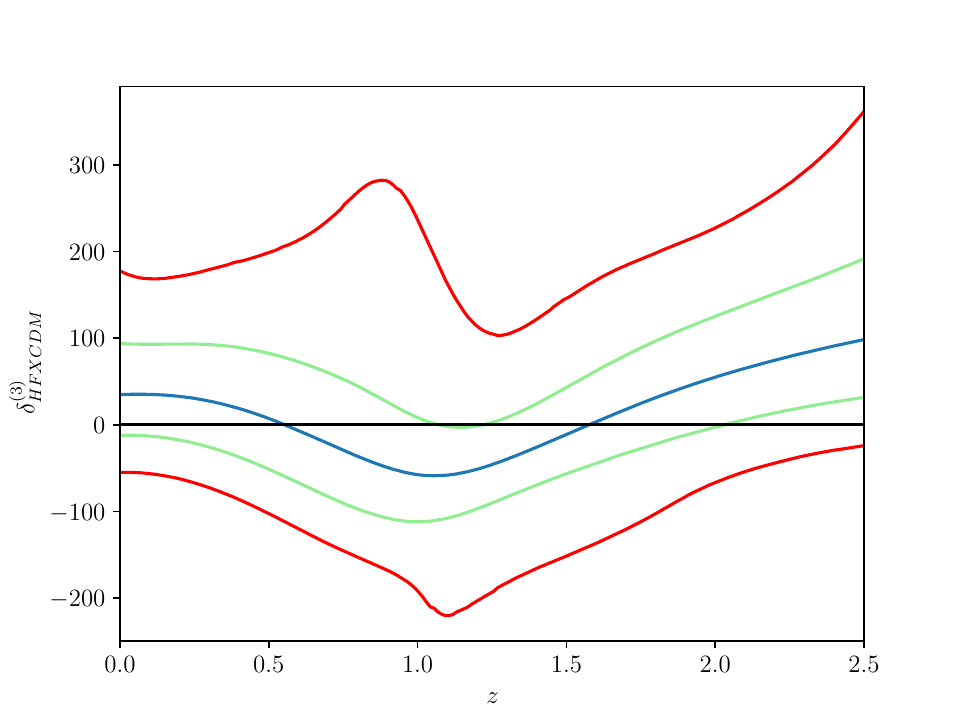}
     \caption{Plot of $\delta^{(3)}_{HF\text{XCDM}}=H'''_{GP}-H'''_{F\text{XCDM}}$, the difference between the Hubble parameter third derivative obtained from GP reconstruction and the same quantity obtained from the flat XCDM relation, as explained on the text.}
     \label{pdFXCDM}
\end{figure}

For the XCDM model, the $\delta^{(3)}_{H{F \text{XCDM}}}$ is given by \eqref{DHFXCDM} and its reconstruction is shown in Fig. \ref{pdFXCDM}. The reconstruction is zero compatible up to 2$\sigma$ uncertainty over the entire redshift range.

\newpage

\subsection{Consistency tests for SNe Ia data}
Using the reconstruction of the transverse comoving distance $D_M(z)$ and its derivatives, obtained by the GP with the Pantheon+ sample data, we have performed consistency tests for the analyzed models.

\begin{figure}[!h]
    \centering
    \includegraphics[width=0.8\textwidth]{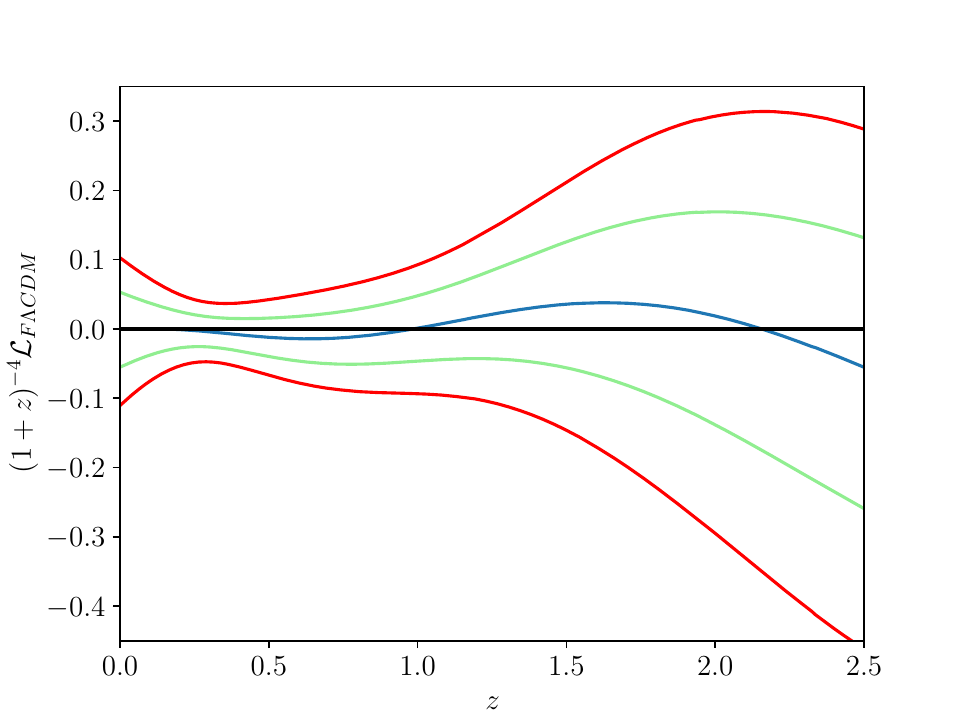}
     \caption{Plot of $(1+z)^{-4}\mathcal{L}_{F\Lambda\text{CDM}}$. As explained on the text, this quantity should be compatible with zero for a consistent flat $\Lambda$ model.}
     \label{figLFLCDM}
\end{figure}

For the flat $\Lambda$CDM model, we reconstruct the $\mathcal{L}_{F\Lambda \text{CDM}}$, which is given by \eqref{LFLCDM}, the reconstruction is shown in Fig. \ref{figLFLCDM}. Similarly to Ref. \cite{Yahya:2013xma}, we have chosen to plot $(1+z)^{-4}\mathcal{L}_{F\Lambda \text{CDM}}(z)$, ensuring stability of the errors. As explained in \cite{Yahya:2013xma}, we can do this because we are testing the equality of these quantities with zero. As a result, the error bands of the reconstructions do not increase with redshift, as expected. The reconstruction is compatible with zero in less than 1$\sigma$ c.l. across the analyzed redshift range.

\begin{figure}[!h]
    \centering
    \includegraphics[width=0.8\textwidth]{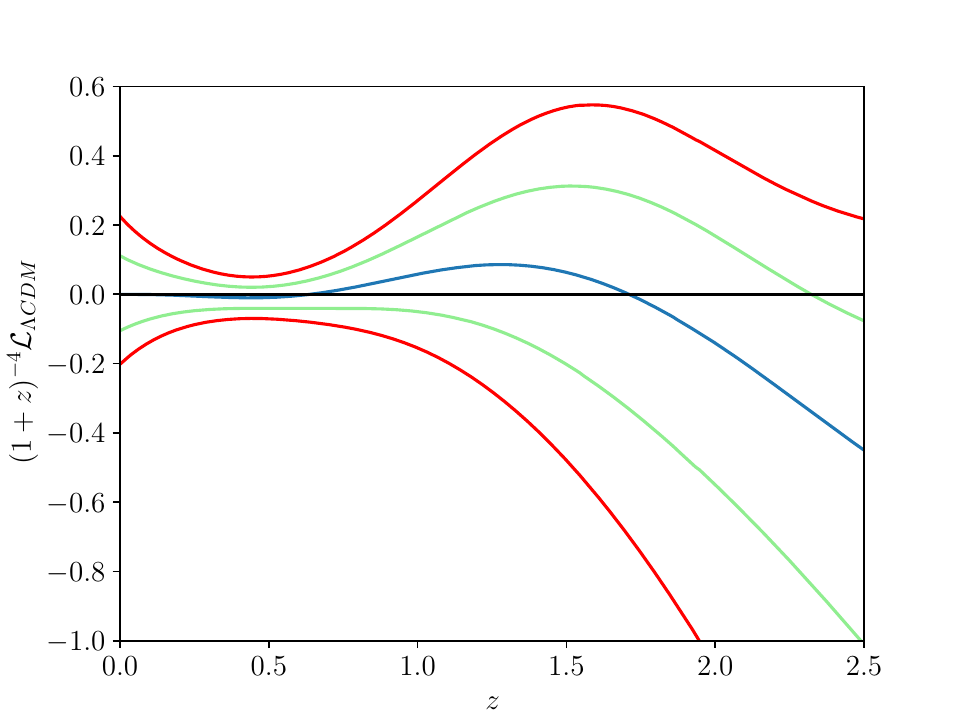}
     \caption{Plot of $(1+z)^{-4}\mathcal{L}_{\Lambda\text{CDM}}$. As explained on the text, this quantity should be compatible with zero for a consistent nonflat $\Lambda$ model.}
     \label{figLLCDM}
\end{figure}

The consistency relation test for the $\Lambda$CDM model was performed using the $\mathcal{L}_{\Lambda\text{CDM}}$, expressed in \eqref{LLCDM}, the reconstruction obtained is shown in Fig. \ref{figLLCDM}. As above, we have chosen to plot $(1+z)^{-4}\mathcal{L}_{\Lambda\text{CDM}}(z)$, ensuring stability of the errors. The reconstruction is compatible with zero in less than 1$\sigma$ c.l. in approximately $93\%$ of the analyzed redshift range.

The consistency tests for the XCDM model were not presented due to divergences in the reconstruction of the consistency relations, as the solutions for XCDM could not be obtained analytically and numerical methods would be necessary.


\section{\label{conclusion}Conclusion}
In the context of $H(z)$ data with $H(z)$ reconstruction, we were able to develop a general method to build consistency relations for cosmological models, and have applied this to some of them, namely, flat and nonflat $\Lambda$CDM and flat XCDM. The analysis of these consistency relations with the current $H(z)$ data, has shown some inconsistency only for flat $\Lambda$CDM in the interval $1.8\lesssim z\lesssim2.4$, at 2$\sigma$ c.l., while nonflat $\Lambda$CDM and flat $\Lambda$CDM were consistent through all the $0<z<2.5$ interval, at this c.l.

Concerning the SNe Ia data with comoving distance reconstruction, we were able to obtain specific consistency relations for specific cosmological models, namely, flat and nonflat $\Lambda$CDM. The analysis of these consistency relations with the current SNe Ia data, namely, the Pantheon+ compilation, has shown that the flat $\Lambda$CDM model was consistent throughout the whole $0<z<2.5$ interval, at 1$\sigma$ c.l., while the nonflat $\Lambda$CDM model was consistent in the same interval, but at only 2$\sigma$ c.l.

A possible way to extend the present results is through new $H(z)$ and SNe Ia data, which can give rise to more stringent reconstructions, thereby showing more inconsistencies of the models or confirming them. Therefore, in this line, new surveys are essential in order to confirm or show any inconsistencies in the context of this analysis.

The methods reviewed and developed here can also be applied to other cosmological models. However, when the model is more complex, the equations involved are yet more complex, and possibly one has to appeal to non-analytical, numerical methods.

\begin{acknowledgments}
RV is supported by  Funda\c{c}\~ao de Amparo \`a Pesquisa do Estado de S\~ao Paulo - FAPESP (thematic projects process no. 13/26258-2, no. 21/01089-1 and regular project process no. 16/09831-0). SHP acknowledges the National Council for Scientific and Technological Development (CNPq) under grants 303583/2018-5 and 308469/2021-6. This study was financed in part by the Coordena\c{c}\~ao de Aperfei\c{c}oamento de Pessoal de N\'ivel Superior - Brasil (CAPES) - Finance Code 001.
\end{acknowledgments}


\end{document}